\documentclass[10pt,letterpaper,twocolumn,amsmath, amssymb]{article}
\usepackage{opex3}
\usepackage[draft]{hyperref}%[draft][colorlinks=true, citecolor=black=, linkcolor=black, urlcolor=blue]
\usepackage{amsmath}
\usepackage{graphicx,dcolumn,bm,color,bbding,epstopdf}
\usepackage{tabularx}
\usepackage{float}
\usepackage[labelsep=period, justification=justified, margin={0.5in,0.5in}, font=footnotesize, singlelinecheck=true]{caption}
\usepackage{cancel}
\usepackage{cite}
\usepackage{rotating}
\usepackage{makecell}

\begin{document}
	%---------------------------
\onecolumn 
\title{Photon pair generation from compact silicon microring resonators using microwatt-level pump powers}

\author{Marc Savanier,$^{1,*}$ Ranjeet Kumar,$^1$ and Shayan Mookherjea$^{1}$}

\address
{
$^1$Department of Electrical \& Computer Engineering, University of California, San Diego,\\ La Jolla, California 92093, USA
%$^*$Corresponding author: \href{mailto:msavanier@ucsd.edu}{msavanier@ucsd.edu}
}
\email{$^*$ msavanier@ucsd.edu}

\begin{abstract}
Microring resonators made from silicon are becoming a popular microscale device format for generating photon pairs at telecommunications wavelengths at room temperature. In compact devices with a footprint less than $5\times 10^{-4}\ \text{mm}^2$, we demonstrate pair generation using only a few microwatts of average pump power. We discuss the role played by important parameters such as the loss, group-velocity dispersion and the ring-waveguide coupling coefficient in finding the optimum operating point for silicon microring pair generation. Silicon photonics can be fabricated using deep ultraviolet lithography wafer-scale fabrication processes, which is scalable and cost-effective. Such small devices and low pump power requirements, and the side-coupled waveguide geometry which uses an integrated waveguide, could be beneficial for future scaled-up architectures where many pair-generation devices are required on the same chip.
%with many pair-generation devices on the same chip, which will be required to create quasi-deterministic pure single photon sources from inherently statistical processes such as spontaneous four-wave mixing.  
\end{abstract}

 
%----------------------

\bibliographystyle{osajnl}
%\bibliography{../../../BibTex_files/ringrefs,../../../BibTex_files/Franson,../../../BibTex_files/Quantum,../../../BibTex_files/CROW_refs}

%-------------------------------------------------------------
\section{Introduction}
Photon pairs and heralded single photons at room temperature are usually generated in nonlinear crystals such as periodically-poled KTP or $\text{LiNbO}_3$ using a non-deterministic nonlinear optical process (spontaneous parametric down-conversion, SPDC). Such devices use a waveguide structure, which is typically a few centimeters long, and require several milliwatts of optical pump power, for which the photons generated through such processes can be contaminated with higher-order photon components\cite{Mosley2008}. For example, a few dozen multiplexed non-deterministic sources with heralding detectors, used concurrently with an active optical switching network has the potential to operate as a quasi-deterministic single photon source with high emission probability \cite{Bonneau2015}.
%A few dozen non-deterministic pair sources used concurrently and with heralding detectors has the potential to operate as a quasi-deterministic pure single photon source with high emission probability\cite{christ2012limits}.
Hence, it is essential to reduce the pump power requirements and investigate compact device structures which can be easily fabricated on a single, stable platform.  

To reduce the device size and the pump power requirements from the milliwatt to the microwatt regime, researchers have been studying resonators, such as the sphere, toroid, disc, ring and photonic crystal resonators\cite{art-Vahala-2003}. There are both fundamental and practical advantages to using small resonators for pair generation. The rate of nonlinear optical processes increases as the mode volume decreases\cite{bk-Sakoda}. Also, if the goal is to generate photon pairs at a single pair of wavelengths, the larger free-spectral range (FSR) makes it easier to extract the signal/idler photons from the ``comb-like'' multiplexed state where many wavelength-pairs are
simultaneously generated\cite{Chen2011}.
%Also, the free-spectral range (FSR) increases, which makes it easier to extract a single pair of wavelengths by filtering. If the goal is to generate photon pairs at a single pair of wavelengths, the larger FSR of small resonators uses the pump power much more efficiently, compared to a dense ``comb-like'' multiplexed state where many wavelength-pairs are simultaneously generated from the shared pump power\cite{Chen2011}.

\begin{figure}[ht]
\centerline{\includegraphics[width=\linewidth]{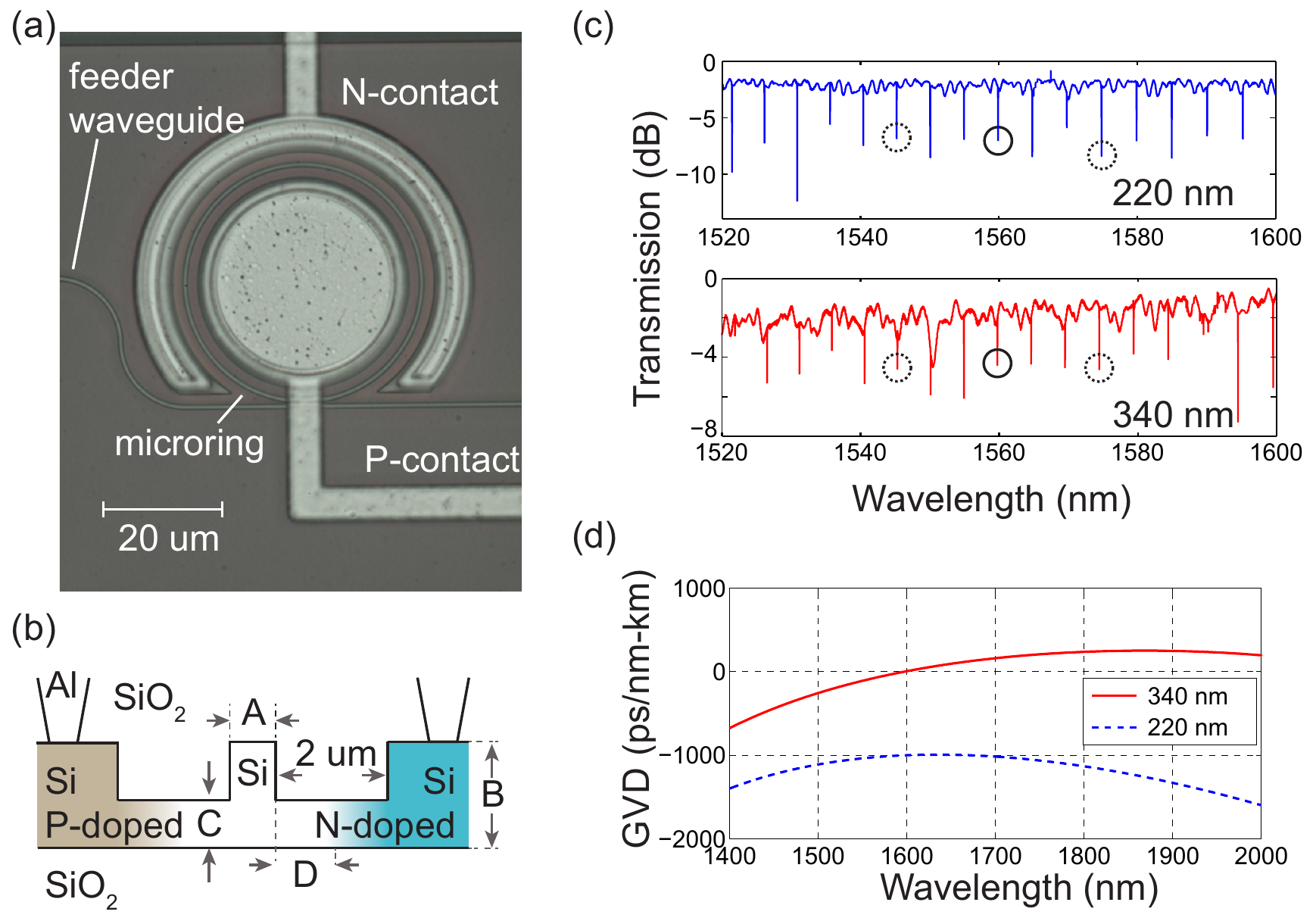}}
\caption{(a) Microscope plan view of a microring resonator with embedded $p$-$i$-$n$ junction across the waveguide. (b) Cross-section schematic drawing; measured devices have $A = 650\ \mathrm{nm}$, $B = 220\ \mathrm{nm}$ or $B = 340\ \mathrm{nm}$, $C = 70\ \mathrm{nm}$, $D = 900\ \mathrm{nm}$. (c) Transmission spectrum for $B = 220\ \mathrm{nm}$ and $B = 340\ \mathrm{nm}$ microring resonators (the solid circles indicate the chosen pump wavelengths, and the dotted circles the measured signal-idler resonances). (d) Group velocity dispersion (calculation) for the lowest-order TE modes of waveguides with the cross-section shown in panel (b).} 
\label{fig-device}
\end{figure}

Furthermore, if the micro-resonator can be made small enough (i.e., with a diameter of approximately $10-50\ \mathrm{\mu m}$, as shown in Fig.~\ref{fig-device}), then the FSR is moderately large (several nanometers), and maps conveniently into the devices operating around the 1550~nm wavelength that have already been designed for telecommunications. Fiber-pigtailed micro-optical components, including reconfigurable add/drop filters, wavelength-division multiplexers / de-multiplexers, etc., are inexpensive and improve the practicality of the overall system compared to breadboard-sized assemblies used in conventional pair-generation experiments using nonlinear crystals.    

Since forming such small-footprint micro-resonators using KTP and $\text{LiNbO}_3$ is difficult (although recent progress has been reported\cite{wang2014integrated}), researchers have studied silicon photonics in which small micro-resonators can be easily made. The physical mechanism for photon pair generation is spontaneous four-wave mixing (SFWM), which is a $\chi^{(3)}$-based nonlinear optical process\cite{LinAgrawal2006}, compared to $\chi^{(2)}$-based SPDC in crystals. The highest quality ($Q$) factors of silicon micro-resonators are found in the toroid and undercut disc geometries, with $Q >10^6$ having been reported\cite{borselli2005beyond,Jiang:15}, but the fabrication and coupling techniques are difficult to scale up.  In contrast, microrings can be more easily fabricated, and can also be integrated more easily with waveguides and other integrated optical structures. In previous work, hundreds of silicon microrings were coupled for classical optics (e.g., ``slow light'' delay lines or for optical filtering)\cite{Cooper2010}, and upto 35 coupled microrings were used for photon pair generation\cite{davanco2012}. 

For these reasons, we focus our study on single waveguide-coupled silicon microring resonators (fabricated on a chip that contains many other functional components, including high-performance filters\cite{6527974} and wavelength converters\cite{art-Ong-FWM-2013}). Here, we demonstrate higher $Q$-factors for silicon microring resonators used for pair generation than reported so far (see Table~\ref{tab-filter}), and discuss the role of design parameters such as loss, group-velocity dispersion and ring-waveguide coupling coefficient, making progress towards room-temperature pair generation that uses microwatt-level optical pumps. 

%---------------------
%---------------------

\section{Fabrication}
Our devices were fabricated at IME (Singapore) using deep ultraviolet lithography on 200~mm silicon-on-insulator (SOI) wafers with $3\ \mathrm{\mu m}$ buried oxide thickness. The waveguide cross-section is shown schematically in Fig.~\ref{fig-device}(b). Two etch steps were used to give a rib waveguide cross-section of height (labeled `B' in Fig.~\ref{fig-device}) 220 nm for one batch of microrings, and a height of 340 nm for another batch. In both cases, the waveguide width (labeled `A' in Fig.~\ref{fig-device}) was 650 nm, and the slab thickness (labeled `C') was 70 nm. The reported transmission spectra in Fig.~\ref{fig-device}(c) correspond to $R$ = 20 $\mu$m radius microrings with inter-waveguide gap of 200 nm (satisfying the minimum feature-size recommendation for the foundry fabrication process). Only the lowest-order transverse electric (TE) polarized mode family propagates in such microrings. Consequently, each transmission spectrum shows a clean, single-family set of resonance peaks as shown in Fig.~\ref{fig-device}(c).

 The $p$ and $n$ doped regions were formed by implanting Boron and Phosphorous, respectively, with the edge of the implanted regions being 900~nm from the edge of the ridge (labeled `D' in Fig.~\ref{fig-device}). These implanted regions create a photodiode which monitors the optical power in the microring, so that the pump wavelength can be aligned to the microring resonance despite temperature variations, as discussed elsewhere\cite{savanier2015optimizing}.  

\section{Microring resonators for photon pair generation}
The process of SFWM in silicon photonics\cite{Sharping2006,LinAgrawal2006}, uses a nonlinear optical interaction seeded by a single-wavelength (continuous-wave or pulsed) pump to generate a pair of photons at two different wavelengths which are prescribed by energy conservation and phase matching. This work focuses on the common type-0 SFWM where the interacting photons are co-polarized, which is the preferred configuration when using ridge Si waveguides. Attempting to use cross-polarized SFWM with, e.g., square cross-sectional waveguides may substantially increase the propagation loss, and lower the quality factors from those achieved here.

In resonators, a frequency ``comb'' consisting of multiple pairs of wavelengths is generated\cite{Chen2011,kumar2013spectrally}. Earlier papers have discussed the physics for SFWM in microrings\cite{Chen2011,Helt:12} but the derived expressions for pair-generation rate (PGR) in \cite{Chen2011,Helt:12} require modifications. In particular, \cite[Eq.~(6)]{Chen2011} focuses on the signal-idler spectra only, without including the power enhancement on resonance in the PGR prefactor, and \cite[Eq.~(62)]{Helt:12} does not have the usual squared-sinc term for phase-matching, which appears in the expression for SFWM pair generation in waveguides\cite{Xiong2011}. The expression for PGR in the case of resonant SFWM in \cite{Helt:12} was derived from first-principles time-domain coupled mode theory when \{pump,~signal,~idler\} resonant triplets are perfectly phase-matched. More generally, we derive an expression for PGR which includes the phase-matching term explicitly. This is important because the phase-matching condition in materials with a high nonlinear coefficient depends also on the optical power, not just the device chromatic dispersion.

In the traveling-mode picture, due to the circulation of the fields in a microring resonator, nonlinear optical interactions between modes coherently build up intensities over an increased interaction length, which is, intuitively, the resonator length multiplied by the number of round trips made by resonant photons (dictated by the `photon lifetime' in the resonator). As discussed in the Appendix, we used the transfer matrix model of propagation for a waveguide-coupled microring resonator \cite{art-Yariv-EL} to ``unfold'' the resonator path into an equivalent conventional waveguide, and show that both the interaction length and the circulating powers are scaled by the resonator finesse $F$.  Thus, referring to treatments of SFWM in waveguides which retain the phase-matching terms explicitly \cite{Xiong2011}, we derived an expression for the the PGR of a microring resonator $r$ under continuous-wave pumping:
\begin{equation}
r = \Delta\nu\left[\gamma\,P^{res}(\lambda)\,L_{\mathrm{eff}}^{res}\right]^2\;\mathrm{sinc}^2\left(\beta_2\,\Delta\omega^2\frac{L^{res}}{2}+\gamma\,P^{res}\,L_{\mathrm{eff}}^{\mathrm{res}}\right)
	\label{eq-PGR}
\end{equation}
where $\Delta\nu=c/\lambda_PQ$ is the ring resonance width, $\gamma$ is the waveguide nonlinearity, $\Delta\omega$ is the pump--signal(idler) angular frequency separation, and the superscript $res$ indicates the resonantly enhanced quantities following:

\begin{equation}
\begin{gathered}
P^{res}(\lambda) = P\times\frac{F}{\pi}\times\frac{\left(\lambda_P/2Q\right)^2}{\left(\lambda-\lambda_P\right)^2+\left(\lambda_P/2Q\right)^2}\\
L^{res} = L\times \frac{F}{\pi}\\
L_{\mathrm{eff}}^{res} = \frac{1-e^{-\alpha L}}{\alpha}\times \frac{F}{\pi}
\end{gathered}
\end{equation}
with $F=Q\,\lambda_{P}/n_{g} L$, $n_{g}$ and $\alpha$ the group index and propagation loss of the ring resonator constituent waveguide at wavelength $\lambda_{P}$ (length $L = 2\pi R$). We assumed that the monochromatic pump wavelength $\lambda$ is positioned close to the ring resonance wavelength $\lambda_P$. (Aside: due to the thermo-optic effect, $\lambda_P$ may depend on the pump power, and high pump power levels may cause unstable pair generation. At moderate pump powers, the detuning of the resonant wavelength from that of the ``cold'' cavity can be monitored and compensated\cite{savanier2015optimizing}.)  In our experiments, the optical filters used to pick off and detect the signal and idler photons have a bandwidth that is larger than of the microring resonances. The ring resonator quality factor $Q$ combines the contributions of two loss mechanisms -- propagation loss and coupling to the bus waveguide. The loaded quality factor of a ring resonator side-coupled to a bus waveguide reads \cite{BookHeebner}:
\begin{subequations}
\label{Qloaded}
\begin{align}
Q_{L} &=\ \frac{\pi}{2\arcsin\left(\frac{1-a\tau}{2\sqrt{a\tau}}\right)}\,\frac{n_{g}L}{\lambda}\label{Qloaded-a}\\
	 &\underset{a\tau\rightarrow1}{\approx}\ \frac{\pi\sqrt{a\tau}}{1-a\tau}\,\frac{n_{g}L}{\lambda}\label{Qloaded-b}
\end{align}
\end{subequations}
where $a = \exp(-\alpha L/2)$ is the round-trip field attenuation, $n_{g}$ is the group index, and $\tau^2 = 1 - \vert\kappa\vert^2$ is the (path-integrated) coupler intensity transmission coefficient, with $\vert\kappa\vert^2$ being the path-integrated coupling coefficient which appears in the popular matrix formulation of the waveguide-ring model \cite{art-Yariv-EL}. The approximation made in Eq.~(\ref{Qloaded-b}) is valid in the high-$Q$ regime towards which all recent results are progressing (see Table~\ref{tab-filter}). For our devices, it holds within 10\% upto $\vert\kappa\vert^2$ = 0.92.

For an isolated resonator ($\tau$ = 1), the unloaded $Q$ is dominated by loss and represents the intrinsic $Q$ limit. In the low-loss regime, it can be expressed as:
\begin{equation}
\label{Qloss}
Q_{U} = \frac{2\pi n_{g}}{\lambda\alpha}.
\end{equation}
In the case where the loaded quality factor is dominated by the coupling coefficient in the low-$\vert\kappa\vert^2$ limit, $Q$ is given by:
\begin{equation}
\label{Qcoupl}
Q_{\mathrm{cpl}} = \frac{2\pi n_{g}L}{\lambda\vert\kappa\vert^2}.
\end{equation}

Note that $n_{g}$, $\alpha$, and the GVD coefficient ($\beta_2$) depend on the waveguide cross-section, and that $\vert\kappa\vert^2$ depends on the gap between the feeder waveguide and the microring.
%\begin{equation}
%Q^{-1} = Q_{\mathrm{loss}}^{-1} + Q_{\mathrm{cpl}}^{-1} = \left(\frac{2\pi\,n_{g}}{\lambda\alpha}\right)^{-1} + \left(\frac{2\pi\,n_{g}L}{\lambda|\kappa|^2}\right)^{-1}.
%\end{equation}
%
%Note that both the group index ($n_{g}$), and the loss ($\alpha$), and the GVD ($\beta_2$) coefficients depend on the waveguide cross-section, and that the intensity coupling coefficient ($|\kappa|^2$, a path-integrated coupling coefficient which appears in the matrix-formulation of the waveguide-ring model\cite{art-Yariv-EL}) depends on the gap between the feeder waveguide and the microring.
 
\subsection{Basic parameters}
The ``nanowire'' silicon waveguide, with a cross-section that is approximately 525~nm wide and 226~nm tall\cite{Dulkeith:06}, supports a single optical mode in the TE and TM polarizations, and the latter can be excluded from participating in the transmission resonances not only by injecting the optical input in the TE polarization, but also by relying on the higher bend loss of the TM mode compared to the TE mode\cite{Vlasov:04}. However, making a waveguide narrow and short enough to exclude all other modes results in high group-velocity dispersion (4400 ps/nm-km)\cite{Dulkeith:06}, and incurs a significant optical propagation loss (typically 2--4 dB/cm)\cite{Vlasov:04,6324087}, both of which can negatively impact pair generation.

Our waveguide cross-sections, shown in Fig.~\ref{fig-device}(b) have a more favorable GVD coefficient, calculated to be -1040 ps/nm-km ($B$ = 220 nm) or -110 ps/nm-km ($B$ = 340 nm) at a wavelength of 1550 nm. Test waveguides similar to those used to form the microring resonator were measured using an atomic force microscope to have a root-mean-squared sidewall roughness of 2.6~nm, and, using an optical cutback method, to have a propagation loss of -0.74 dB/cm ($B$ = 220~nm) and -1.23 dB/cm ($B$ = 340~nm) at 1550~nm.
% The calculated dispersion times length factor $(\beta_2L)^{1/2} is \approx 0.09\ \mathrm{ps}$ for a 220~nm microring with effective propagation length $L = \tau c/n_g \approx 6.1\ \mathrm{mm}$, based on the group index $n_g$. Based on the measured linewidth of the resonance, the photon lifetime $\tau$ was estimated to be $79\ \mathrm{ps}$. 

The resonant enhancement in a resonator can be quantified through its quality factor ($Q$), and the standard microring theory \cite{Niehusmann:04} can be used to estimate an upper-bound on the intrinsic (unloaded, loss-limited) $Q$'s, if we assume that the loss of the microring-waveguide directional coupler is negligible. These are estimated as $Q_U = 9.2 \times 10^5$ ($B$ = 220 nm) and $Q_U = 5.6 \times 10^5$ ($B$ = 340 nm). From the transmission measurement performed at low input power to prevent thermo-optic effect, our fabricated microring resonators were seen to have a (loaded) quality factor around $1 \times 10^5$ e.g., $B$ = 220~nm microring with radius $20\ \mathrm{\mu m}$: $Q_L \approx 9.5 \times 10^4$ and $B$ = 340~nm microring with radius $20\ \mathrm{\mu m}$: $Q_L \approx 2.5 \times 10^5$. These $Q$'s compare favorably with other reports (see Table~\ref{tab-filter}).

%---------------------
\begin{figure}[ht]
\centerline{\includegraphics[width=0.5\textwidth]{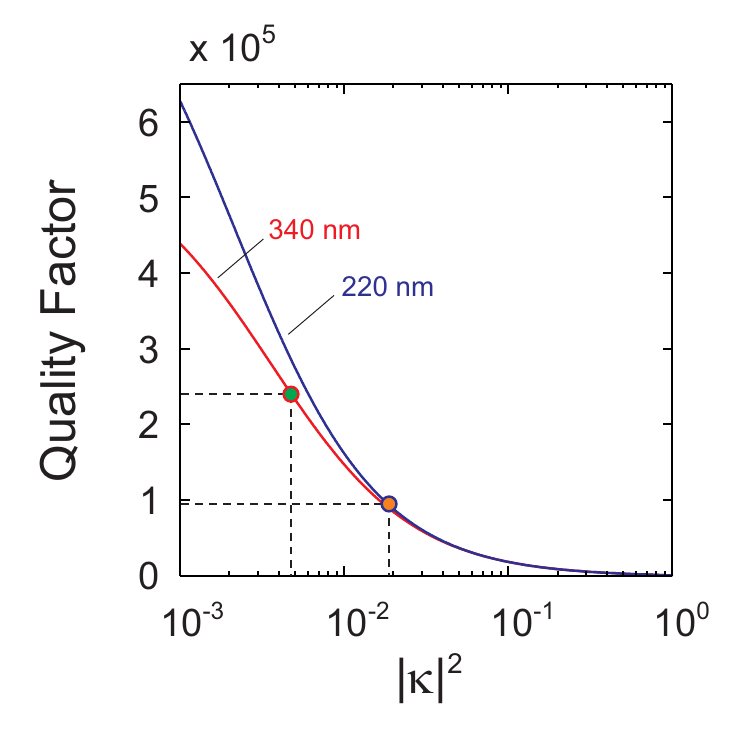}}
\caption{Nonlinear relationship between the quality factor $Q$ and the coupling coefficient $\vert\kappa\vert^2$, calculated using Eq.~(\ref{Qloaded-a}) and the parameters specific to the two structures studied here ($B$ = 220 nm: blue, $B$ = 340 nm: red). The $\vert\kappa\vert^2$ values for the fabricated devices are estimated from the measured loaded $Q$'s (dots and dashed lines)}
\label{FNew}
\end{figure}
%---------------------

Equation~(\ref{Qloaded}) shows that $Q_L$, the experimentally accessible parameter, depends on the propagation loss $\alpha$ and the coupling coefficient $\vert\kappa\vert^2$. While $\alpha$ should be as low as possible, $\vert\kappa\vert^2$ may be selected by the experimenter during the design stage. Figure~\ref{FNew} makes explicit the correspondence between $Q_L$ and $\vert\kappa\vert^2$ for the microring devices presented above, based on the measured value of $\alpha$ and the simulated $n_g$ (Lumerical software package). From $Q_L = 9.5 \times 10^4$ ($B$ = 220~nm) and $Q_L = 2.5 \times 10^5$ ($B$ = 340~nm), we inferred the respective coupling coefficients $\vert\kappa\vert^2 \approx 0.018$ and $\vert\kappa\vert^2 \approx 0.005$.

\subsection{Microring-waveguide coupling coefficient}
The value of $|\kappa|^2$, the path-integrated coupling coefficient of the waveguide directional coupler formed between the feeder waveguide and the microring resonator\cite{art-Yariv-EL}, is determined by the width of the waveguides, the separation between them, and the length of the coupling region \cite{SoltaniCoupling}. The benefit of using deep ultraviolet lithography though a well-calibrated silicon foundry process is that it is easier to accurately fabricate directional couplers with a small desired value of $|\kappa|^2$, as is required for a high-$Q$ resonator. 

\begin{figure}[ht]
\centerline{\includegraphics[width=0.5\textwidth]{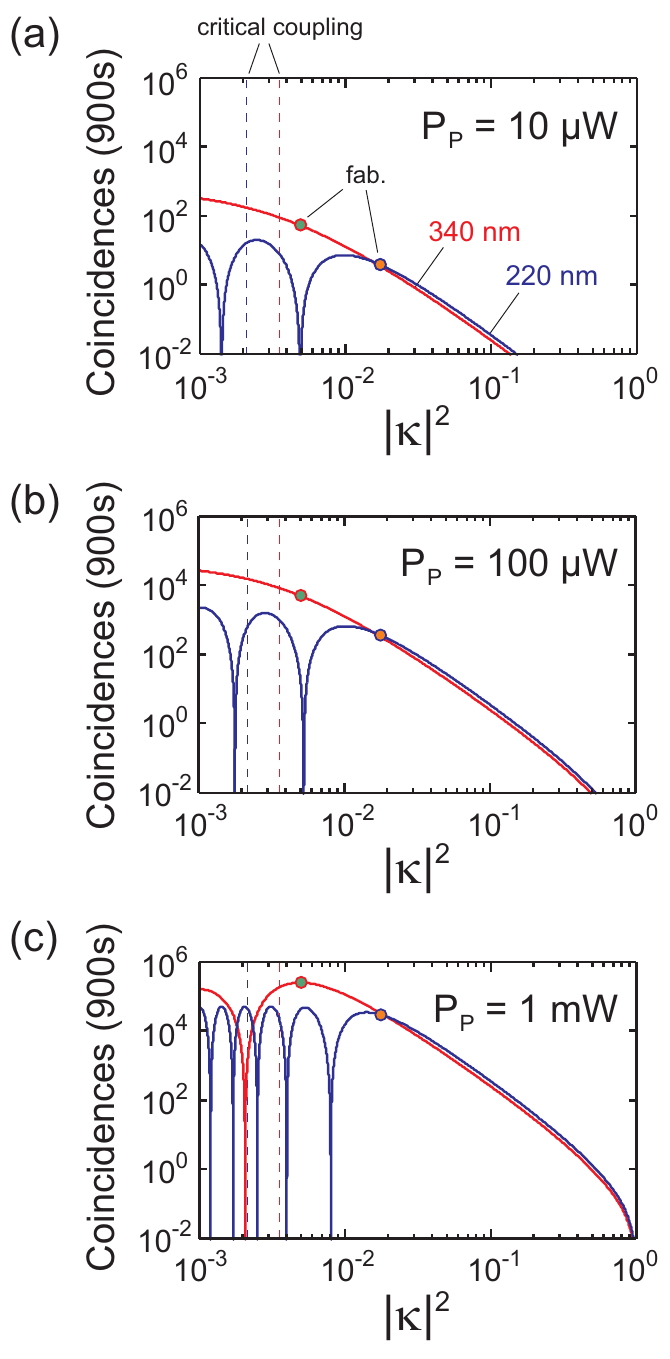}}
\caption{The appropriate choice of the coupling coefficient $|\kappa|^2$ is guided by numerical calculations of the coincidences counted in 900~s (with assumed values of the fiber-to-chip coupling efficiency, filter insertion loss and detector efficiency that simulate the experiments performed in this report). A study of two cross-sections ($B$ = 220~nm and 340~nm -- blue and red curves respectively) for increasing pump powers: (a) 10~$\mu$W, (b) 100~$\mu$W and (c) 1~mW, shows the critical role of GVD and self-modulation on the coincidences count, through the SFWM phase-mismatch.}
\label{F2}
\end{figure}

Figure~\ref{F2} shows calculations of the PGR using Eq.~(\ref{eq-PGR}) for different values of $|\kappa|^2$. The horizontal axis is plotted on a logarithmic scale, since we are especially interested in small values of $|\kappa|^2$. Since the phase-matching `sinc' term in Eq.~(\ref{eq-PGR}) depends on optical power, we have shown the power-dependent behavior of PGR in a sequence of three representative cases. For these numerical calculations, we have used values for the fiber-waveguide coupling efficiencies, waveguide losses and detector parameters which are relevant to our experiment, but altering these values does not qualitatively change the following observations. 

%---------------------
%---------------------

As shown by Fig.~\ref{F2}, PGR does not grow indefinitely for small values of $|\kappa|^2$, as may be supposed from an idealized model\cite{Helt:12}. This is because the `sinc' term in Eq.~(\ref{eq-PGR}) decreases in magnitude if either $P$ or $L$ increases. These behaviors are important once the $Q$ factor of the microring exceeds a reasonably high value --\,around 10$^5$ for typical GVD of Si waveguides, see Fig.~\ref{FNew}\,-- and may not have been relevant for lower-$Q$ microring resonators (see Table~\ref{tab-filter}). The PGR scales as the cube of $Q$ and in the search to improve PGR by achieving higher $Q$ through improved fabrication, we expect these effects will become more important. (Another effect, non-linear absorption, which has been studied elsewhere\cite{Guo:14}, would also limit the increase of PGR at high $Q$ values.) For small values of $|\kappa|^2$ (i.e., under-coupled regime), the optical propagation length increases, and there are `nulls' in the PGR (driven by the `sinc' term). Practically, it may be safer to use larger values of $|\kappa|^2$ (i.e., the slightly over-coupled regime) to reduce the risk that minute fabrication imperfections--or even temperature-dependent fluctuations--would catastrophically lower the PGR. For these slightly larger values of $|\kappa|^2$, to the right-hand side of the optimum in Fig.~\ref{F2}, there are no nulls, since the argument of the `sinc' term is small. The circles indicate the values of $|\kappa|^2$ used in our devices, following this principle. These nulls are obviously much more of a factor if the pump power is increased.  The benefit of the lower GVD of the $B$ = 340~nm waveguide (red curve) is evident, compared to the 220~nm tall waveguide (blue curve).

We may choose to design (or tune) the microring for critical coupling at the pump wavelength, as shown by the dashed lines in Fig.~\ref{F2}. At first glance, this can be beneficial because the maximum input optical pump power is delivered to the resonator. From a practical perspective, the transmission of the residual pump power past the microring is minimized at critical coupling. Any assistance in extinguishing the pump that can be provided by critical coupling helps filtering, since on-chip silicon photonic filters do not provide as much on-off contrast as off-chip assemblies. (There are a few reports \cite{6527974,harris2014integrated} of approximately 100~dB filtering contrast using silicon integrated optics, and one direct measurement of 100~dB contrast wavelength scan \cite{6527974}. Using multiple chips involves fiber interconnections, making the assembly comparable to simply using off-the-shelf telecommunications fiber-optic components to perform the filtering. At present, bulk optical filtering assemblies designed for SPDC still demonstrate higher performance \cite{hockel2010note}, and it would be beneficial if comparable performance could be achieved on a single microchip platform.)
%\footnote{There are a few reports of approximately 100~dB filtering contrast using silicon integrated optics, including our previous work\cite{6527974,harris2014integrated}; however, \textit{all\/} these reports actually use combinations of multiple chips, or estimate the performance of multiple stages from adding the contrast of single stages (which is not conclusive). No single-chip measurement of a filter contrast exceeding about 70~dB has been reported. An actual measurement of 100~dB contrast in Ref. \cite{6527974} is, as far as we know, unique. Using multiple chips involves fiber interconnections, making the assembly no more convenient than simply using off-the-shelf telecommunications fiber-optic components to perform the filtering. In contrast, bulk optical filtering assemblies designed for SPDC currently demonstrate much higher performance\cite{hockel2010note}.}.
In silicon microrings formed using low-GVD waveguides, the value of $|\kappa|^2$ that achieves critical coupling is, in fact, close to the value of $|\kappa|^2$ which optimizes PGR at low pump powers. However, the strong variations in PGR from the phase-matching term near critical coupling (for the range of pump powers shown in Fig.~\ref{F2}) do not provide much tolerance against fabrication imperfections, compared to the moderately over-coupled regime. Furthermore, should a microring resonator be operated as a heralded single photon source, using a picosecond-scale short-pulse pump laser, the optimum heralding rate is also found at a moderately over-coupled regime \cite{VernonHerald}.

As shown by Fig.~\ref{F2}, our experimentally-fabricated couplers (shown by the circles) achieved values of $|\kappa|^2$ that were close to, and slightly above, critical coupling (shown by the dashed red and blue lines). We performed three-dimensional finite-difference time-domain (FDTD) method simulations of the directional coupler region using the Lumerical software package, from which we inferred that the difference between our fabricated couplers and the critical coupling condition corresponded to an alteration in defining the ring-waveguide gap distance by 43~nm for the $B$ = 220~nm ring, and by 16~nm for the $B$ = 340~nm ring. (Note that typical tunable directional couplers used in silicon photonics based on electro-optic or thermo-optic Mach-Zehnder interferometers are too long to be used in such compact microrings under an FSR constraint.)

\section{Measurements}
The microresonator was optically pumped in the usual experimental setup for SFWM which we have described previously\cite{davanco2012}. Based on a separate calibration measurement, we measured the fiber-to-waveguide insertion loss to be -3~dB. Photons at the output of the chip were spectrally separated using a custom-built tunable filter assembly (using telecommunications components) that suppressed the residual pump photons by more than 150~dB and extracted the signal and idler wavelengths, with insertion loss -6~dB and -5~dB, respectively. Single photons at the signal and idler wavelengths were detected using gated thermo-electrically cooled (234~K) InGaAs SPADs with an estimated 15$\%$ quantum efficiency, electrically-generated gate width of 2.5~ns and gating repetition frequency 5~MHz \cite{Tosi-2012}. The dark counts of the two SPADs were 70~Hz and 130~Hz.  

For a representative microring in these experiments ($B$ = 220~nm, $R = 20\ \mathrm{\mu m}$), the intensity enhancement factor (ratio of the circulating field intensity in the microring to the intensity in the feeder waveguide) was 173. Thus, in the low pump power regime, an input waveguide power of -10~dBm resulted in a circulating field intensity in the microring of about $9.3\ \mathrm{MW}.\mathrm{cm}^{-2}$. The associated two-photon absorption in silicon was calculated to be 0.02~dB/cm, i.e.~more than one order of magnitude smaller than the reported linear propagation loss. Pair generation was stabilized by finding the wavelength at which the pump transmission was minimized, and ``soft-locking'' the pump wavelength to the transmission minima\cite{Carmon:04}. 

%---------------------
\begin{figure}[ht]
\centerline{\includegraphics[width=0.5 \columnwidth]{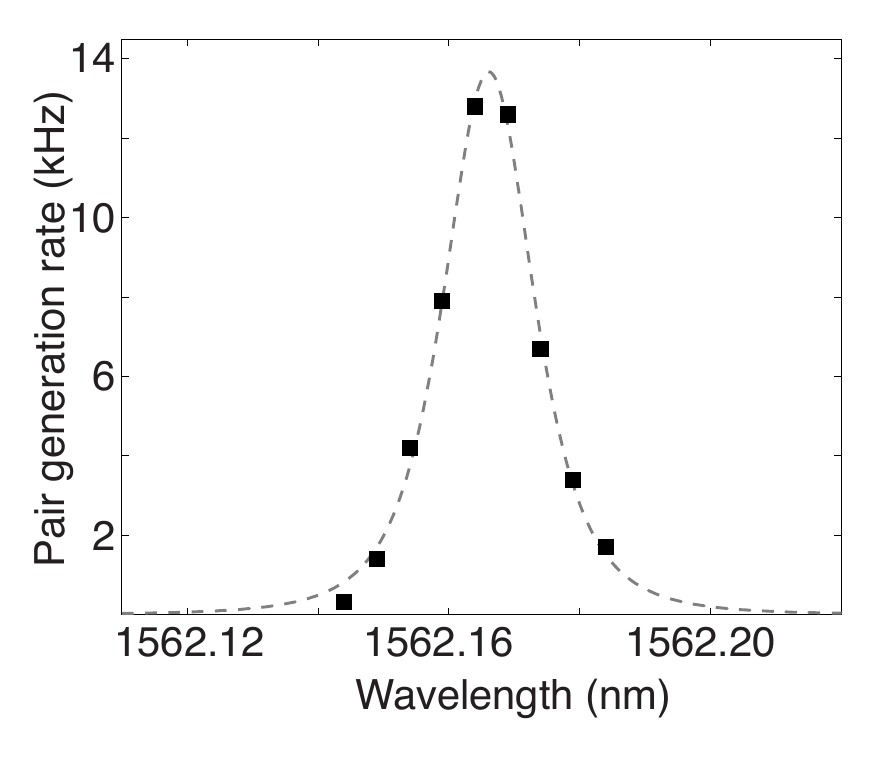}}
\caption{Measurements of PGR as the pump wavelength was finely tuned across the resonance (the error bars are smaller than the markers); input pump power was approximately -10 dBm. The dashed line is a squared-Lorentzian fit.} 
\label{fig-Lorentzian}
\end{figure}
%---------------------

As a function of the pump wavelength, for un-filtered (i.e., integrated over their respective resonance) signal-idler photons, the PGR peaks when the pump is spectrally aligned to the resonant wavelength of the microring. (Although this appears intuitive, such data has not yet been reported.) Indeed, Fig.~\ref{fig-Lorentzian} shows measurements (continuous-wave pump, gated SPADs--width 2.5~ns, rep. rate 50~MHz) of the PGR made by changing the pump wavelength in small steps across a resonance of a microring ($B$~=~220~nm, $R = 10\ \mathrm{\mu m}$). This measurement has been performed at low input pump power for which self-phase modulation can be neglected, resulting in a constant value of the `sinc' function in Eq.~(\ref{eq-PGR}). Following Eq.~(\ref{eq-PGR}), the dashed line shows a good fit by a squared-Lorentzian functional form, since the field generated by SFWM is proportional to the square of the circulating pump field, and the latter may be assumed to follow the Lorentzian line shape. 

Under these operating conditions, at a temperature of 30$^\circ$C and input pump power (in the feeder waveguide) of -11~dBm at wavelength $\lambda_{P} =$ 1562.16~nm, the signal and idler photon pairs were generated at $\lambda_{S} =$ 1542.61~nm and $\lambda_{I} =$ 1582.24~nm, respectively with a generation probability $\approx$ 2 $\times$ 10$^{-4}$ pairs/detector gate.

%---------------------
\begin{figure}[ht]
\centerline{\includegraphics[width=0.70\columnwidth]{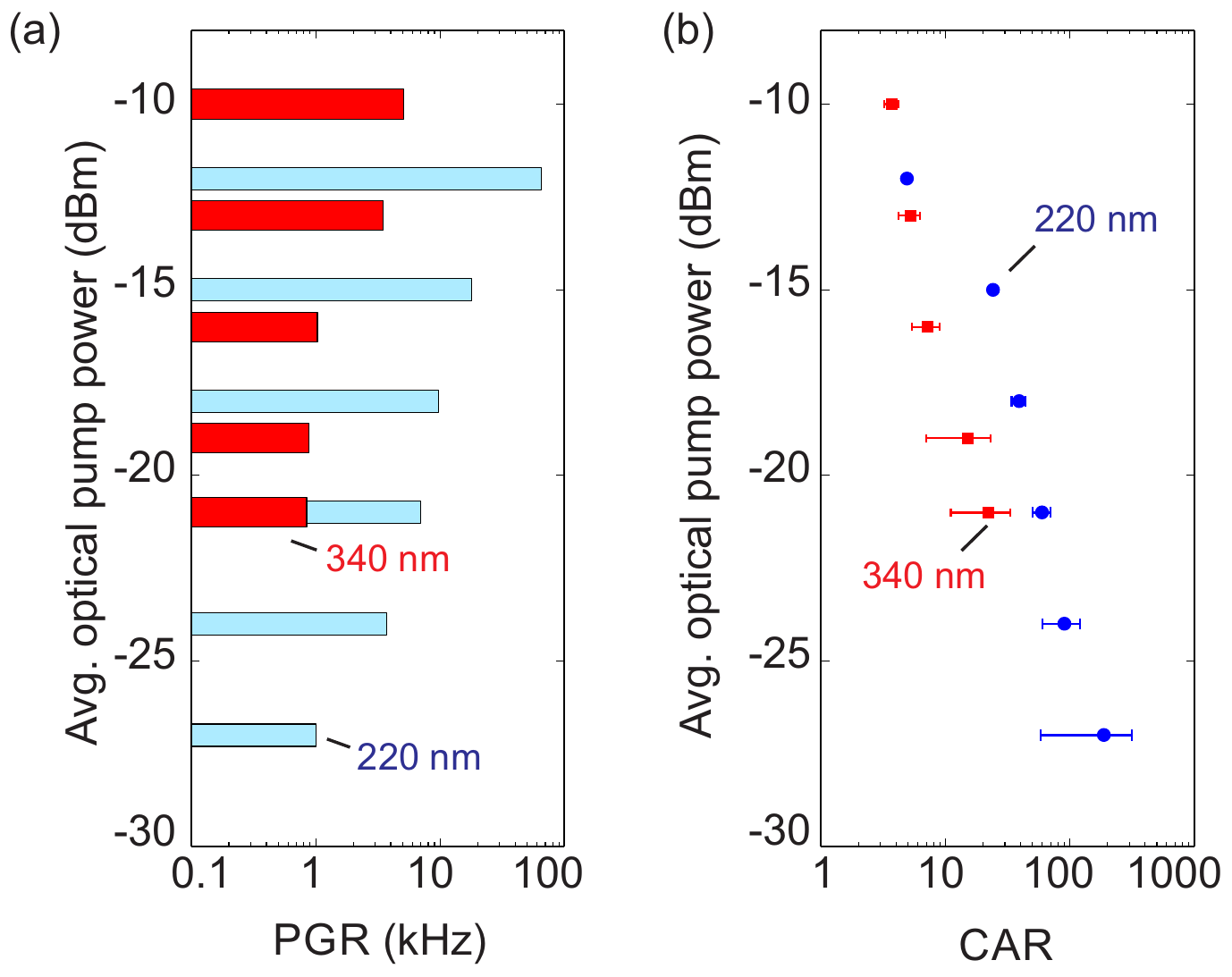}}
\caption{(a) Measured pair generation rate (PGR) using pump pulses of 3~ns duration at a repetition rate of 5~MHz (the error bars are too small to be visible) and (b) coincidences-to-accidentals ratio (CAR) versus average optical pump power in the feeder waveguide.} 
\label{fig-PGR-CAR}
\end{figure}
%---------------------

Figure~\ref{fig-PGR-CAR}(a) shows the PGR measurements of $R = 20\ \mathrm{\mu m}$ radii devices for $B$~=~220~nm and $B$~=~340~nm. The experimental PGR are calculated from the measured coincidence rates by accounting for the chip-to-fiber coupling loss, the off-chip filter losses, and detector quantum efficiencies, for which numerical values were provided in the earlier paragraphs. We further scaled the PGR by the duty cycle of the SPADs to calculate the on-chip ``intrinsic'' PGR, which we reported alongside other results of silicon microring pair generation in Table~\ref{tab-filter}. The coincidences-to-accidentals ratio (CAR), plotted in Fig.~\ref{fig-PGR-CAR}(b), is an often-used figure-of-merit used to characterize the generation and detection of photon pairs. The improvement in CAR at lower pump powers (and thus, at lower PGR) has been reported and discussed extensively in previous papers\cite{Sharping2006,Harada2008,clemmen2009continuous,Engin:13,Grassani:15}. Having a signal at least one order of magnitude stronger than noise (i.e., CAR \textgreater\,10) suggests a possible implementation for applications such as quantum cryptography and quantum key distribution (QKD). While there is no hard rule for CAR with an application, CAR $\sim$ 10 has been used previously in the literature, e.g., \cite{Takesue-CAR} explicitly uses this value to numerically estimate an 5\% error rate of a QKD system using entangled photon pairs and time correlation measurements.

%-------------------------------------------------------------
\begin{table}
\caption[table]{Recent results of photon pair generation using silicon microring resonators (\textit{comparison: recent microdisk result})}
\label{tab-filter}
{\raggedleft
\begin{tabularx}{0.87\textwidth}{|c|c|c|c} \hline\hline
Ref. & Dimensions$^{(a)}$ & Radius & Quality Factor ($Q$) \\ \hline\hline
Clemmen et al.\cite{clemmen2009continuous} & $500\times 200\ \mathrm{nm}$ & $6.8\ \mathrm{\mu m}$ & $4.5 \times 10^4$\\
Azzini et al.\cite{Azzini2012} & $500\times 220\ \mathrm{nm}$ & $5\ \mathrm{\mu m}$ & $7.90 \times 10^3$ \\
Engin et al.\cite{Engin:13} & $450\times 220\times 50\ \mathrm{nm}$ & $11\ \mathrm{\mu m}$ & $3.75 \times 10^4$ \\
Harris et al.\cite{harris2014integrated} & $500\times 220\ \mathrm{nm}$ & $15\ \mathrm{\mu m}$ & $4 \times 10^4$ \\
Silverstone et al.\cite{Silverstone:2015} & $500\times 220\ \mathrm{nm}$ & $15\ \mathrm{\mu m}$ & $9.2 \times 10^3$ \\
Guo et al.\cite{Guo:14} & $450\times 220\times 60\ \mathrm{nm}$ & $21\ \mathrm{\mu m}$ & $8.1 \times 10^4$ \\
Grassani et al.\cite{Grassani:15} & $500\times 220\ \mathrm{nm}$ & $10\ \mathrm{\mu m}$ & $1.5 \times 10^4$ \\
Wakabayashi et al.\cite{Wakabayashi2015} & $400\times 220\ \mathrm{nm}$ & $7\ \mathrm{\mu m}$ & $2 \times 10^4$ \\
Gentry et al.\cite{Gentry:2015} & $1080\times \textrm{sub-}100\ \mathrm{nm}$ & $22\ \mathrm{\mu m}$ & $3.1 \times 10^4$ \\
This work & $650\times 340\times 70\ \mathrm{nm}$ & $20\ \mathrm{\mu m}$ & $2.5 \times 10^5$ \\
This work & $650\times 220\times 70\ \mathrm{nm}$ & $20\ \mathrm{\mu m}$ & $9.5 \times 10^4$ \\
This work & $650\times 220\times 70\ \mathrm{nm}$ & $10\ \mathrm{\mu m}$ & $9.6 \times 10^4$ \\ \hline
\itshape Jiang et al.\cite{Jiang:15} & \itshape [Microdisk] & $\mathit{5\ \mu m}$ & $\mathit{5 \times 10^5}$ \\ \hline
\end{tabularx}}
\newline
{\center
\begin{tabularx}{\textwidth}{cX|c|c|c|c|} \hline\hline
\textit{Continued}& & GVD & PGR & Average pump power & CAR\\ \hline\hline
Clemmen et al.\cite{clemmen2009continuous}& & -0.7 $\text{ps}^2.\text{m}^{-1}$ & 300 kHz & $400\ \mathrm{\mu W}$ & 30\\
Azzini et al.\cite{Azzini2012}& & - & 200 kHz & $200\ \mathrm{\mu W}$ & 250 \\
Engin et al.\cite{Engin:13}& & - & 123 MHz & $4.8\ \mathrm{mW}$ & 37$^{(b)}$ \\
Harris et al.\cite{harris2014integrated}& & - & 600 kHz & $300\ \mathrm{\mu W}$ & 50 \\
Silverstone et al.\cite{Silverstone:2015}& & - & 4.6 MHz & $150\ \mathrm{\mu W}^{(b)}$ & 10 \\
Guo et al.\cite{Guo:14}& & - & 14 kHz & $100\ \mathrm{\mu W}$ & 180 \\
Grassani et al.\cite{Grassani:15}& & - & 10 MHz & $1\ \mathrm{mW}$ & 65 \\
Wakabayashi et al.\cite{Wakabayashi2015}& & - & 21 MHz & $410\ \mathrm{\mu W}$ & 352 \\
Gentry et al.\cite{Gentry:2015}& & - & \makecell{165 Hz\\29 kHz} & \makecell{$5\ \mathrm{\mu W}$\\$50\ \mathrm{\mu W}$} & \makecell{37\\55} \\
This work& & +0.14 $\text{ps}^2.\text{m}^{-1}$& 57 kHz & $8\ \mathrm{\mu W}^{(c)}$ & 22 \\ 
This work& & +1.33 $\text{ps}^2.\text{m}^{-1}$& 470 kHz & $8\ \mathrm{\mu W}^{(c)}$ & 60\\ 
This work& & +1.33 $\text{ps}^2.\text{m}^{-1}$& 83 kHz & $79\ \mathrm{\mu W}$ & 65\\ \hline
\itshape Jiang et al.\cite{Jiang:15}& & - & $\mathit{855\ kHz}$ &  $\mathit{79\ {\mu W}}$ & $\mathit{274}$\\ \hline 
\end{tabularx}}
\newline $^{(a)}$ With references to Fig.~1(b): $A\times B \times C$.
\, $^{(b)}$ Peak power = 0.25~W. 
\, $^{(c)}$ Peak power = $530\ \mathrm{\mu W}$. 
\end{table}
%-------------------------------------------------------------

 %-------------------------------------------------------------
\section{Discussion}
\begin{figure}[ht]
\centerline{\includegraphics[width=0.8\columnwidth]{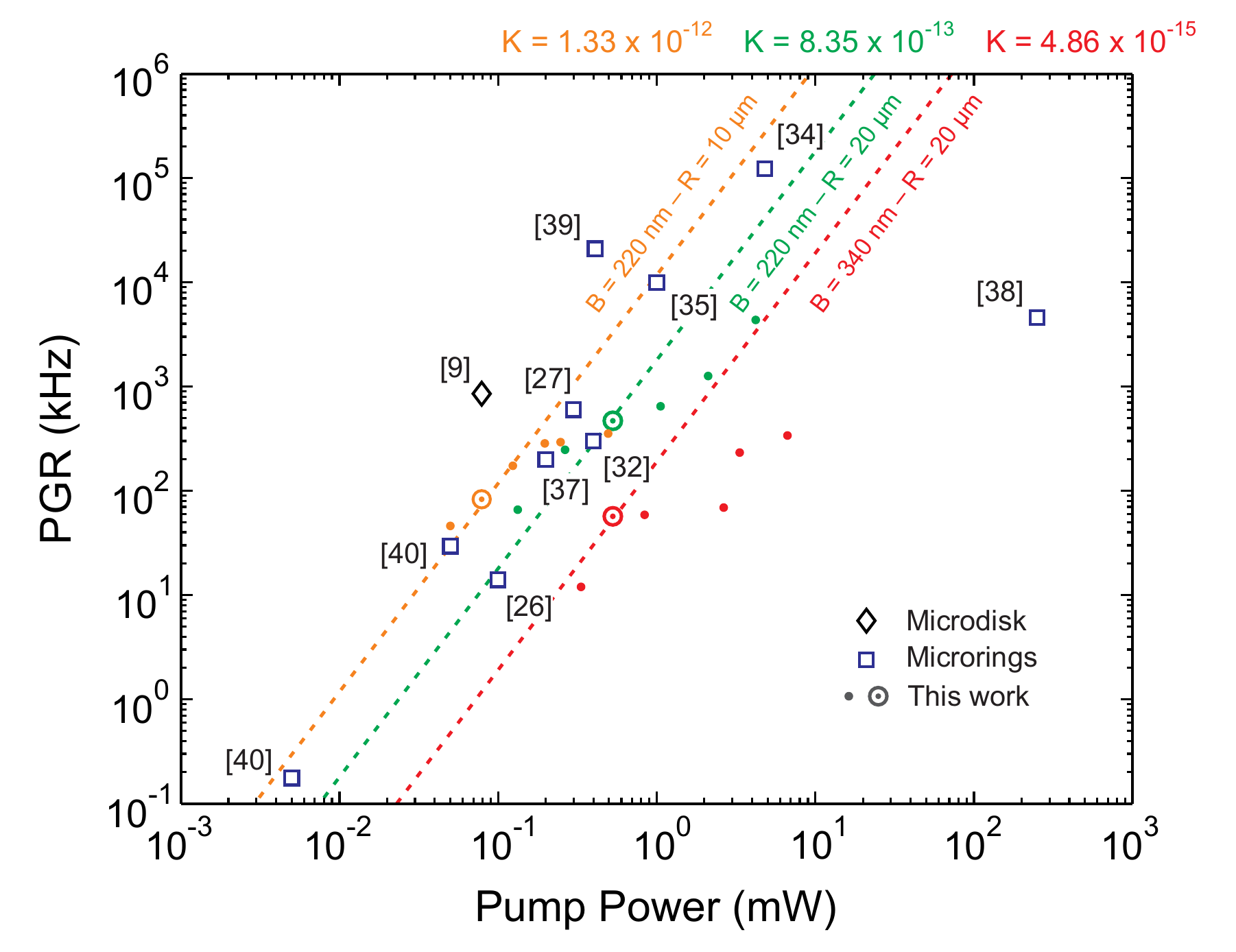}}
\caption{Measured on-chip pair generation rate (PGR) versus average optical pump power ($P$) for the devices presented here (dots), along with the recent experimental reports of Table~\ref{tab-filter} in microrings (squares), and microdisk (diamond). The quadratic dependence of PGR with $P$ is highlighted by the fits $PGR = KQ^3/R^2\times P^2$ of each experimental dataset at low pump power (dashed lines), while saturation occurs at higher pump powers.}
\label{fig-PGR-Pp}
\end{figure}

The scaling of PGR versus $P$ is shown in Fig.~\ref{fig-PGR-Pp}, with additional details about the experimental parameters provided in Table~\ref{tab-filter}. Generally, microring resonators with smaller radii are more efficient at generating pairs, as has already been discussed in \cite{Azzini2012}.

%---------------------
%--------------------- 

However, these PGR values are still several orders-of-magnitude below what theory predicts. In the ideal case where the SFWM process is perfectly phase-matched (i.e., `sinc' is unitary),  we calculated that the two microring resonators studied here would result in PGR = 364~MHz ($B$ = 340~nm) and PGR = 33~MHz ($B$ = 220~nm) at $P = 530~\mu$W. One possible reason could be the dependence of the coincidence count on the precise value of $|\kappa|^2$, as discussed earlier.
%However, these PGR values are still several orders-of-magnitude below what has been predicted by theory\cite{Clemmen2010}. One possible reason could be the dependence of the coincidence count on the precise value of $|\kappa|^2$, as discussed earlier. In the ideal case where the SFWM process is perfectly phase-matched (i.e., `sinc' is unitary),  we calculated that the two microring resonators studied here would result in PGR = 3.6~GHz ($B$ = 340~nm) and PGR = 329~MHz ($B$ = 220~nm) at $P = 530~\mu$W. 

Our values of $|\kappa|^2$ (marked `fab.' in Fig.~\ref{F2}) were estimated from the measured linewidth of the resonance, assuming that the waveguide propagation loss coefficient which was measured on separate test structures (paperclip waveguides) also represents the loss coefficient of the (curved) waveguides in the microring, and there are no additional losses (e.g., in the coupler). Changing these assumptions would vary the estimated value of $|\kappa|^2$. Thus, there is some small uncertainty in separating the two factors which determine the ``loaded'' $Q$-factor (propagation loss and $|\kappa|^2$). However, if we assume that our estimation of $|\kappa|^2$ is, in fact, exactly accurate, then Eq.~(1) predicts that our microrings should exhibit PGR = 268~MHz ($B$ = 340~nm) and PGR = 20~MHz ($B$ = 220~nm). These PGR values are already smaller than the ideal, phase-matched cases which ignore the `sinc' term in Eq.~(1). Moreover, as Fig.~\ref{F2} shows, small errors in fabrication can further reduce PGR catastrophically if we happen to fall into one of the nulls of the argument, which are closely spaced at higher pump powers.  

Thus, in practice, the PGR depends critically on the microring-waveguide coupling coefficient, especially when GVD is non-zero and low $|\kappa|^2$ values are targeted. Since fabrication of silicon photonic microrings is already a fairly mature technology, this may be a reason why the experimentally achieved PGR results shown in Table~\ref{tab-filter} are still many orders of magnitude lower than theory\cite{clemmen2009continuous}. Reducing the pump power mitigates these impairments somewhat, but also results in a less bright pair source.  

%There are some additional associated subtleties when using high-$Q$ resonators. As Fig.~\ref{fig-device}(c) shows, the depth of the transmission nulls changes significantly and quasi-randomly from one resonance to another (and from one device to another, although the measurements are entirely repeatable). This is a fairly common observation when measuring high-$Q$ resonators. Consequently, the fraction of the input pump power that is coupled into the microring depends on which resonances are used. For example, the ratio of the pump powers coupled from the waveguide into the ring in the $B$ = 220~nm case (which has the deeper transmission resonance) to the $B$ = 340~nm case was 1.70, and the square of this factor (i.e.,~$2.9$) should be used in comparing the PGR of the two microrings at similar input (waveguide) power levels.

Based on Eq.~(\ref{eq-PGR}) in the low-loss approximation, we also fitted the experimental data at low optical power following $PGR = KQ^3/R^2\times P^2$. The quadratic behavior was seen to hold for in-coupled pump power up to $\approx100~\mu$W ($R$ = 10~$\mu$m) and $\approx500~\mu$W ($R$ = 20~$\mu$m), above which the PGR saturated. In general, pair generation in microrings operating at high pump power is hampered by nonlinear effects triggered by high circulating power (e.g.,~two-photon absorption, free-carrier absorption/dispersion, thermal shift). 

The structure of these microrings (see Fig.~\ref{fig-device}) shows a $p$-$i$-$n$ junction diode fabricated across the waveguide ridge. Originally, this was done for the purposes of sweeping out optically-generated free-carriers and thus improving the PGR, following the report of \cite{Engin:13}. However, despite careful study, we were unable to prove this effect in any of our large ensemble of chips. Instead, the diode performed a different, and very useful, purpose of providing an electronic readout (essentially, a Ge-free photodiode) for aligning the pump laser to the microring, as we reported recently\cite{savanier2015optimizing}. Those measurements showed that the optimal pump wavelength for peak PGR was the same as the wavelength for minimum pump transmission past the microring.

 %-------------------------------------------------------------
\section{Conclusion}
Table~1 summarizes the advancement in the state-of-the-art in photon pair generation using silicon microring resonators at wavelengths near $1.55\ \mathrm{\mu m}$. With these few-microwatt pump power pair-generation results, silicon microring resonators now approach a performance regime previously only attainable using microdisk resonators\cite{Jiang:15}, compared to which microrings have some advantages such as a cleaner mode spectrum, CMOS-compatible fabrication and waveguide coupling instead of suspended-fiber coupling. Higher $Q$ factors can be achieved than reported here\cite{biberman2012ultralow}, but at the cost of a larger mode area (thus, a smaller nonlinear coefficient) and a much larger bending radius (millimeters rather than microns), which is not desirable for photon pair generation. In this report, we found that compact silicon microrings with a radius of about $10\ \mathrm{\mu m}$ (or less, as long as the loss can be kept low),  which are slightly over-coupled to the feeder waveguide and are operated at low pump power will come closest to the theoretical performance predictions. In fact, since these devices can be pumped with only a few tens of microwatts of average pump power to achieve hundreds-of-kilohertz rate (on-chip) PGR, it may even be possible to use on-chip microlasers to realize a fully-integrated chip-scale source of photon pairs.\\

%-------------------------------------------------------------
\section*{Appendix: Phase Mismatch in a Dispersive Microring Resonator}

\setcounter{equation}{0} \renewcommand{\theequation}{A\arabic{equation}}

A detailed model of nonlinear interactions in the traveling-mode picture in a microring resonator will be presented elsewhere; here, we focus on showing the structure of the linear phase mismatch term in Eq.~(\ref{eq-PGR}) in the context of classical nonlinear four-wave mixing. This has not been shown in previous discussions of microring resonators\cite{BookHeebner,Chen2011,art-Absil-2000}. In particular, we show that the length $L$ that appears in the `sinc' phase-matching term is a resonantly-enhanced length, i.e.,~a multiple of the physical resonator circumference $d = 2 \pi R$ (radius $R$), and that the multiplicative constant is linearly proportional to the finesse $F$. Consequently, a high-finesse (high-$Q$) microring resonator has a tighter phase-matching constraint than a lower-finesse micro-resonator. 

\begin{figure}[ht]
\centerline{\includegraphics[width=0.8\linewidth]{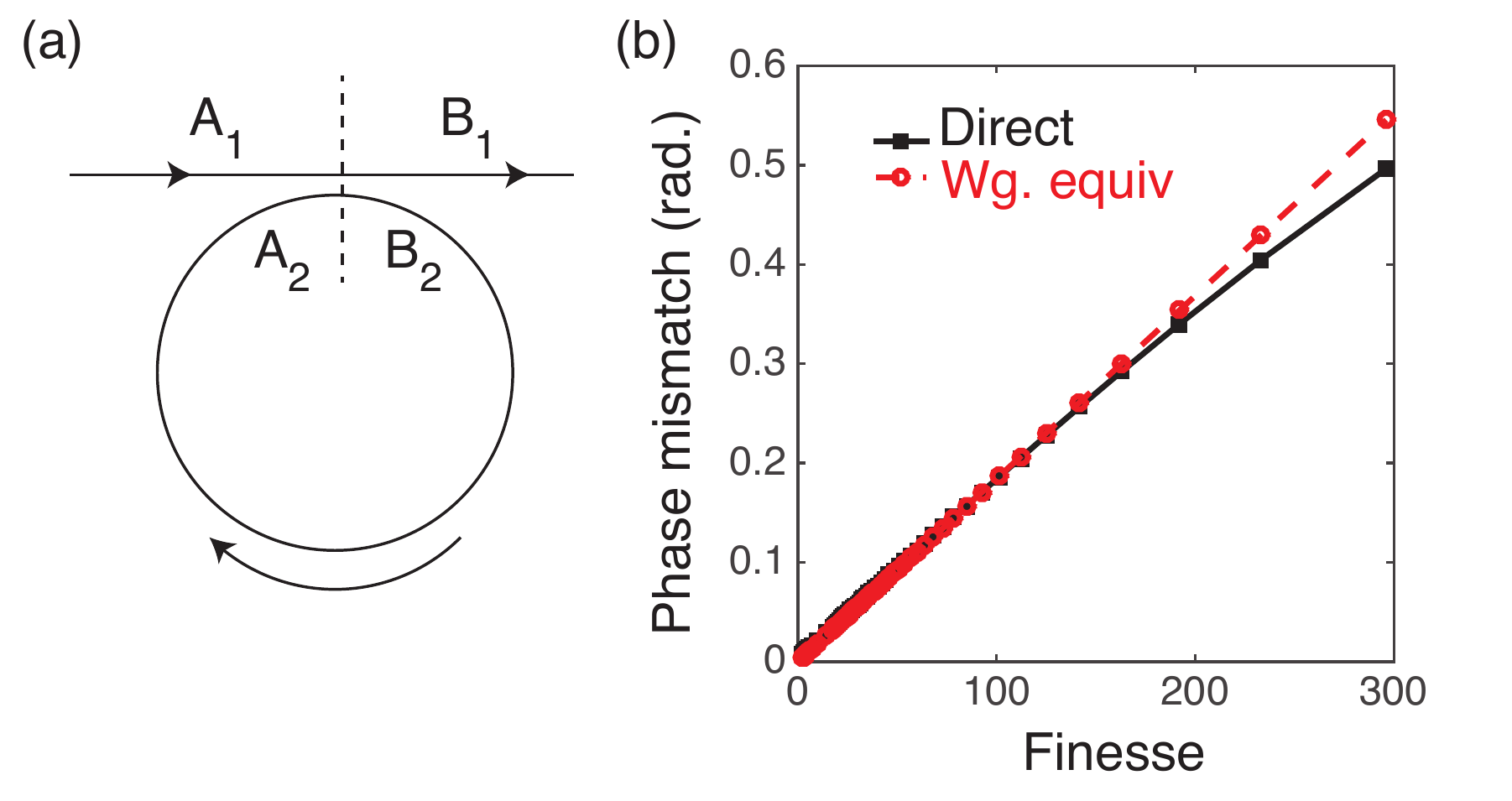}}
\caption{(a) Schematic of microring resonator coupled to a waveguide, with electric field amplitude coefficients which are used in the matrix model of propagation. (b) Calculation of the phase mismatch versus finesse, using two approaches as discussed in the text. } 
\label{fig-App-1}
\end{figure}

Consider the simple model of a microring with a directional (point) coupler to a waveguide (see~Fig.~\ref{fig-App-1}(a) and \cite[Fig.~4.22]{bk-YarivYeh6}), in which optical fields are described by four amplitude coefficients, $A_1$, $A_2$, $B_1$ and $B_2$, where $A_1$ is the input amplitude from the feeder waveguide just before the coupler, $B_1$ is the field amplitude in the waveguide just after the coupler, and $A_2$ and $B_2$ are the field amplitudes in the microring before and after coupler, respectively. Using a matrix formalism and ignoring the self-phase modulation nonlinear phase accumulation terms, we define the (normalized) circulating field in the microring as $H$:
\begin{equation}
H \equiv \frac{A_2}{A_1} = \frac{\kappa a\, e^{i \phi}}{1-\tau a\, e^{i\phi}},
\end{equation}
where $a = \mathrm{exp}(-\alpha d/2)$ is the amplitude change and $\phi=k d$ is the phase accumulated in a single round trip by an optical wave with propagation constant $k$, i.e.,~$A_2 = a e^{i\phi} B_2$, and the directional coupler is described by through- and cross-coupling amplitude coefficients, $\tau$ (real) and $\kappa$ (imaginary), respectively, and the matrix equation:
\begin{equation}
\left( \begin{array}{c} B_1 \\ B_2 \end{array}  \right) = 
\left( \begin{array}{cc} \tau & \kappa^* \\ \kappa & -\tau^* \end{array}  \right)
\left( \begin{array}{c} A_1 \\ A_2 \end{array}  \right).  
\end{equation}
The phase of $H$ is:
\begin{equation}
\Phi \equiv \mathrm{arg}(H) = \frac{\pi}{2} + \phi + \mathrm{tan}^{-1}\left(\frac{\tau a \sin \phi}{1- \tau a \cos \phi} \right).
	\label{eq-phase-def}
\end{equation}
In the vicinity of a resonance ($\phi \approx m\, 2\pi$, where $m$ is an integer), 
\begin{equation}
\Phi \approx \frac{\pi}{2} + \phi + \left(\frac{\tau a}{1- \tau a} \right) \phi = \frac{\pi}{2}+\left(\frac{1}{1-\tau a}\right)\phi.
	\label{eq-phase-def-simple}
\end{equation}
%As $a\tau \rightarrow 1$, the last term in the above expression dominates the others. 

For four-wave mixing in a conventional waveguide of length $L$ near the phase-matched condition, the small phase mismatch between the pump (subscript `p'), signal (`s') and idler (`i') wavelengths is 
\begin{equation}
\Delta k L = 2k_pL - k_sL - k_iL \approx \beta_2 \Delta \omega^2 L \label{eq-phase-mismatch-waveguide}
\end{equation}
where $\Delta \omega$ is the angular frequency separation between the signal and the pump, or between the pump and the idler, and $\beta_2=\frac{d^2k}{d\omega^2}$ is the group velocity dispersion of the waveguide taken at $\omega_p$. In the microring device, we define the phase mismatch, which plays the same role as $\Delta k L$,  as
\begin{equation}
\Delta \Phi = 2\Phi_p - \Phi_s - \Phi_i. \label{equation-phase-mismatch}
\end{equation}
where each $\Phi$ term is calculated using Eq.~(\ref{eq-phase-def}) at the appropriate wavelength. We calculated the wavelength dependence of the modal effective and group indices for our ridge waveguides using the `MODE' software package (Lumerical, Inc.), and, furthermore, assumed that the propagation loss of the waveguide that constitutes the microring was $\alpha = -0.7$~dB/cm.
\vspace{2mm}
%---------------------
%---------------------

In Fig.~\ref{fig-App-1}(b), we choose a range of values of $t$ between $0.31$ and $0.99$, and plot the phase mismatch as a function of the finesse, $F = \pi \left[2\, \sin^{-1}\left(\frac{1-a\tau}{2 \sqrt{a\tau}}\right)\right]^{-1}$, calculated in two ways:

The first method (labeled `Direct') is to (numerically) evaluate $\Delta \Phi$ from Eq.~(\ref{equation-phase-mismatch}). A non-zero phase mismatch arises from the dispersion of the microring's behavior at the pump, signal and idler frequencies, which arises (in our calculation here) from the dispersion of the refractive index. In silicon waveguides, the dispersion is far from zero, and in practice, there may be additional differences because of the dispersion of the directional coupling coefficients, which can be even larger~\cite{art-Aguinaldo-2012}. 

The second approach (`Wg. equiv.') is to conceptually ``unfold'' the microring into an equivalent straight waveguide whose length is equal to the circumference of the microring multiplied by $N = F/\pi$, which is the number of round-trips as dictated by the resonator-enhanced photon lifetime (or equivalently, the quality factor). The phase mismatch is given by Eq.~(\ref{eq-phase-mismatch-waveguide}), and $\Delta k L \approx \beta_2 (\omega_s - \omega_p)^2 (F/\pi) d$, where $\beta_2$ is the waveguide GVD evaluated at $\omega_p$.

The close agreement of the two calculations, except at high finesse values, where the effect of finite waveguide loss is evident in Eq.~(\ref{eq-phase-def}) but not in Eq.~(\ref{eq-phase-mismatch-waveguide}), shows that the ``unfolding'' picture yields the correct phase mismatch. Without this effect, the phase mismatch, e.g., as calculated by the formula presented in~\cite{art-Absil-2000,Chen2011}, is incorrect, by more than two orders of magnitude. The phase-matching bandwidth is reduced in a high-$Q$ resonator by multiplying the ring circumference by the effective number of round-trips that the light propagates, which is equal to the finesse divided by $\pi$. This effect is in addition to the enhancement of the field amplitudes in the microring, which enhances the strength of the nonlinear interaction. These effects are recognized in at least two other related research fields: nonlinear optical interactions in photonic crystal micro-resonators\cite[p.~121]{bk-Sakoda}, and nonlinear interactions in coupled microring resonators\cite{art-Morichetti-2011}, where the language of ``slow light'' (i.e., enhanced group delay on resonance, which is also observed in a single microring resonator coupled to a waveguide) is used.  

We also derived the contribution to phase matching from self-phase modulation, which scales with the square of the finesse, and therefore, results in Eq.~(\ref{eq-PGR}).

%-------------------------------------------------------------
\section*{Acknowledgments}
The authors are grateful to Jun Rong Ong (now at A$^\star$STAR -- IHPC) for device design and discussions, Xianshu Luo and Guo-Qiang Patrick Lo (A$^\star$STAR -- IME) for device fabrication, Nick Bertone (Optoelectronic Components), Alberto Tosi (Politecnico di Milano) and the National Science Foundation for support under grant ECCS 1201308.

\end{document}